\documentclass[12pt]{article}
\renewcommand{\theequation}
{\arabic{section}.\arabic{equation}}

\makeatletter
\def\eqnarray{ \stepcounter{equation} \let\@currentlabel=\theequation
 \global\@eqnswtrue
 \global\@eqcnt\z@
 \tabskip\@centering
 \let\\=\@eqncr
 $$\halign to \displaywidth\bgroup\@eqnsel\hskip\@centering
 $\displaystyle\tabskip\z@{##}$&\global\@eqcnt\@ne
 \hfil$\displaystyle{{}##{}}$\hfil
 &\global\@eqcnt\tw@$\displaystyle\tabskip\z@{##}$\hfil
 \tabskip\@centering&\llap{##}\tabskip\z@\cr}
\makeatother

\makeatletter
\def\@arrayacol{\edef\@preamble{\@preamble \hskip .5\arraycolsep}}
\def\array{\let\@acol\@arrayacol \let\@classz\@arrayclassz
\let\@classiv\@arrayclassiv \let\\\@arraycr\def\@halignto{}\@tabarray}
\makeatother
\makeatletter
\newcounter{subeqncnt}
\def\thesubeqncnt{\alph{subeqncnt}}
\def\subequations{\begingroup%
   \stepcounter{equation}\edef\@tempa{\theequation}%
   \let\c@equation\c@subeqncnt\c@subeqncnt\z@
   \edef\theequation{\@tempa\noexpand\thesubeqncnt}}

\makeatother
\newcommand{\be}{\begin{equation}}
\newcommand{\ee}{\end{equation}}

\newcommand{\beqa}{\begin{eqnarray}}
\newcommand{\eeqa}{\end{eqnarray}}
\newcommand{\nn}{\nonumber}

\newcommand{\eqref}[1]{(\ref{#1})}

\begin{document}

\setlength{\baselineskip}{7mm}
\begin{titlepage}
\begin{flushright}
{\tt NRCPS-HE-54-07} \\
{\tt RH-02-2007} \\
April, 2007
\end{flushright}

\vspace{1cm}

\begin{center}
{\it \Large A Dual Lagrangian for Non-Abelian Tensor Gauge Fields
}

\vspace{1cm}

{ {Jessica K. Barrett $^{a}$}\footnote{jessica(AT)raunvis.hi.is}}
and
{ {George  Savvidy $^{b}$}\footnote{savvidy(AT)inp.demokritos.gr}}

\vspace{1cm}

\item[$^a$]{\it Science Institute,} \\
{\it University of Iceland,}
{\it  Taeknigardi, Dunhaga 5 ,} \\
{\it IS-107 Reykjavik, Iceland} \\

\vspace{0.5cm}

\item[$^b$]{\it Institute of Nuclear Physics,} \\
{\it Demokritos National Research Center }\\
{\it Agia Paraskevi, GR-15310 Athens, Greece}

\end{center}

\vspace{1cm}

\begin{abstract}
For non-Abelian tensor gauge fields of the lower rank we have found
an alternative expression for the field strength tensors, which
transform homogeneously with respect to the complementary
gauge transformations and allow us to construct the dual Lagrangian.
\end{abstract}

\end{titlepage}

%%%%%

\section{\it Introduction }

There are many interesting approaches
to formulating the higher-spin field theories and
tensor gauge field theories.
The Lagrangian and S-matrix formulations of {\it free }
massless Abelian tensor gauge fields
have been  constructed in
 \cite{fierz,fierzpauli,yukawa1,wigner,schwinger,
Weinberg:1964cn,chang,singh,fronsdal,deWit:1979pe,Curtright:1980yk}.
The problem of introducing {\it interactions} appears
to be much more complex and there has been important progress
in defining self-interaction of higher-spin fields
in the light-cone formalism and
in the covariant formulation of the theories
\cite{Bengtsson:1983pd,Bengtsson:1983pg,Bengtsson:1986kh,berends,
Berends:1984wp,fronsdal2,Metsaev:2005ar,Boulanger:2006gr,Francia:2007qt}.
The main idea is to introduce self-interactions using iterations:
starting from the free quadratic Lagrangian for the higher-spin field
one should introduce a cubic, quartic and possibly higher-order
terms to the free Lagrangian
and then check, whether the thus deformed algebra of the initial
group of gauge transformations
still forms a closed algebraic structure in covariant formulation
or whether the Lagrangian
remains Lorentz invariant in the light-cone formalism.

There has been important progress in the development of
interacting field theories in anti-de Sitter space-time background,
which is reviewed in \cite{Bekaert:2005vh,Engquist:2002vr,Sezgin:2001zs}
and is of great importance for the development of string field theory.
It should be noted  that self-interaction of higher spin fields
is naturally generated in string field
theory as well
\cite{Thorn:1985fa,Siegel:1985tw,Witten:1985cc,Siegel:1988yz,Arefeva:1989cp}.
From the point of view of quantum
field theory, string field theory seems to contain an
infinite number of nonrenormalizable interactions,
that is a nonlocal cubic interaction terms that contain an exponential
of a quadratic form in the momenta
\cite{Taylor:2003gn,Taylor:2006ye}.

The concept of local gauge
invariance allows one to define the non-Abelian gauge fields \cite{yang},
to derive their  dynamical field equations
and to develop a universal point of view on matter interactions
as resulting from the exchange of spin-one gauge quanta. Therefore
it is appealing to extend the gauge principle so that it will define the
interaction of gauge fields which carry not only
non-commutative internal charges,
but also arbitrary spins \cite{Savvidy:2005fi}.
For that purpose one should define extended
non-Abelian gauge transformations acting on {\it tensor gauge fields}
and the corresponding field strength tensors,
which will enable the construction of a gauge invariant Lagrangian quadratic in field
strength tensors, as in Yang-Mills theory.
The resulting gauge invariant Lagrangian
defines cubic and quartic self-interactions of charged gauge quanta
carrying a spin larger than one
\cite{Savvidy:2005fi,Savvidy:2005zm,Savvidy:2005ki}.

Here we shall follow the construction described above which is based on the
direct extension of non-Abelian gauge transformations
\cite{Savvidy:2005fi,Savvidy:2005zm,Savvidy:2005ki}.  Recall
that in these publications it was found that there exists
not one but  a {\it pair
of complementary} non-Abelian gauge transformations acting on the same
rank s+1 tensor gauge field
$
A^{a}_{\mu\lambda_1 ... \lambda_{s}}
$.
These sets of gauge
transformations $\delta$ and $\tilde{\delta}$ are defined  in
\cite{Savvidy:2005fi,Savvidy:2005zm,Savvidy:2005ki}. Considering the first
set of gauge transformations $\delta$
one can construct infinite series of forms
$ {{\cal L}}_{s}~( s=1,2,..)$ and ${{\cal L}}^{'}_{s}~( s=2,3,..)$ which are
invariant with respect to the first group
of gauge transformations $\delta$
$$
\delta  {{\cal L}}_{s}=0~~~~~   s=1,2,..~~~~~~~~~~~~~~~~
 \delta {{\cal L}}^{'}_{s}=0~~~~~  s=2,3,..
 $$
and are quadratic in the field strength tensors
$G^{a}_{\mu\nu,\lambda_1 ... \lambda_s}$. This construction of
invariant forms was based on the fact that
field strength tensors $G^{a}_{\mu\nu,\lambda_1 ... \lambda_s}$
transform   homogeneously
with respect to the gauge transformation $\delta$.
Therefore the gauge invariant
Lagrangian describing dynamical tensor gauge bosons of all ranks
has the form \cite{Savvidy:2005fi,Savvidy:2005zm,Savvidy:2005ki,Savvidy:2005at}
$$
{\cal L} = \sum^{\infty}_{s=1}~ g_{s } {{\cal L}}_{s }~+
~ \sum^{\infty}_{s=2}g^{'}_{s } {{\cal L}}^{'}_{s }~.
$$

A natural question which arises in this respect is connected with the
possibility of a similar construction, now for the second group
of complementary gauge transformation
$\tilde{\delta}$. More specifically the question is, can one
construct
"complementary" field strength tensors $\tilde{G}^{a}_{\mu\nu,\lambda_1 ... \lambda_s}$
which transform homogeneously
with respect to the $\tilde{\delta}$? And if yes, then to construct
corresponding invariant forms, the  Lagrangian $\tilde{{\cal L}}$ and to find
a possible relation between the Lagrangians   $\tilde{{\cal L}}$
and ${\cal L}$.

The answer that we found for lower-rank
tensor gauge fields is given in  (\ref{fieldstrengthparticular}) and
(\ref{fieldstrengthparticularextens}).
These new field strength tensors transform homogeneously
(\ref{fieldstrenghparticulartransformation})
with respect to the second group of complementary
gauge transformations $\tilde{\delta}$
and allow us to construct
invariant forms $\tilde{{{\cal L}}}_{2}$ and $\tilde{{{\cal L}}}^{'}_{2}$
presented in (\ref{alternativetotalactiontwo}). Thus we have
two Lagrangian forms ${{\cal L}} $    and
$\tilde{{\cal L}} $
for the same lower-rank tensor gauge fields.
The natural question which arises at this point is to find out a
possible relation between these Lagrangian forms.
We have found  that the dual transformation (\ref{dualitytransformation})
maps $\tilde{{\cal L}} $  into the Lagrangian ${{\cal L}} $.
It is not yet known if this construction of  complementary field
strength tensors $\tilde{G}$ and of the corresponding invariant forms
can be fully extended to higher-rank tensor gauge fields.
In the last section we suggested a possible solution of this problem,
but shall leave
this extension for future studies.

\section{\it Complementary Gauge Transformations}

In the recent papers \cite{Savvidy:2005fi,Savvidy:2005zm,Savvidy:2005ki}
the non-Abelian tensor gauge fields are defined
as rank-$(s+1)$ tensors
\be
A^{a}_{\mu\lambda_1 ... \lambda_{s}}(x),~~~~~s=0,1,2,...
\ee
and are totally symmetric with respect to the
indices $  \lambda_1 ... \lambda_{s}  $. {\it A priori} the tensor fields
have no symmetries with respect to the first index  $\mu$.
This is an essential departure from the
previous considerations (yet see  \cite{Curtright:1980yk}),
in which the higher-rank tensors were totally symmetric
\cite{fierzpauli,schwinger,singh,fronsdal} .
The extended gauge transformation $\delta_{\xi}$
which acts on non-Abelian tensor gauge fields of rank $s+1$
$
A^{a}_{\mu\lambda_1 ... \lambda_{s}}(x),~s=0,1,2,...
$
is defined by the following relations\footnote{The gauge parameters
$\xi^{a}_{\lambda_1\lambda_2...}$ are totally symmetric
tensors. The full transformation is
given in (\ref{polygaugefull}). }:
\beqa\label{polygauge}
\delta_{\xi} A^{a}_{\mu} &=&   \partial_{\mu} \xi^{a} +.... ,~~~~~ \nonumber\\
\delta_{\xi} A^{a}_{\mu\lambda_1} &=&   \partial_{\mu} \xi^{a}_{\lambda_1}+....,\nonumber\\
\delta_{\xi} A^{a}_{\mu\lambda_1 \lambda_2} & =&
\partial_{\mu}\xi^{a}_{\lambda_1\lambda_2} +....\\
........&.&....................................... \nn
\eeqa
The transformations $\delta_{\xi} A^{a}_{\mu\lambda_1 ... \lambda_{s}}(x)$
form  an infinite-dimensional gauge group ${\cal G}$,
on which one can define field strength tensors $G^{a}_{\mu\nu,\lambda_1 ... \lambda_s}$.
The field strength tensors $G^{a}_{\mu\nu,\lambda_1 ... \lambda_s}$
transform homogeneously (\ref{transformationfieldstrenghparticular})
and allow the construction of two infinite
series of gauge invariant forms
$ {{\cal L}}_{s}~( s=1,2,..)$ and ${{\cal L}}^{'}_{s}~( s=2,3,..)$.
These forms are
quadratic in field strength tensors  and the
Lagrangian describing dynamical tensor gauge bosons of all ranks
has the form \cite{Savvidy:2005zm,Savvidy:2005ki}
\be\label{lagrangian}
{{\cal L}} =  {{\cal L}}_1 +  g_2 {{\cal L}}_2 +
g^{'}_2 {{\cal L}}^{'}_2 +...,
\ee
where ${{\cal L}}_1$ is the Yang-Mills Lagrangian.
It had been found that one can select the coupling constants $g_{2}$
and $g^{'}_{2}$ so that the {\it free} part of the Lagrangian
${{\cal L}}=  {{\cal L}}_1 +  g_2({{\cal L}}_2 +  {{\cal L}}^{'}_2) $
exhibits gauge invariance with respect to
enhanced gauge transformations $\tilde{\delta}_{\eta}$ which we shall call
"complementary". It has the following form
\cite{Savvidy:2005ki}:
\beqa\label{freedoublepolygaugesymmetric}
\tilde{\delta}_{\eta} A^{a}_{\mu} &=&  \partial_{\mu}\eta^a +... \nonumber\\
\tilde{\delta}_{\eta} A^{a}_{\mu\lambda_1} &=&   \partial_{\lambda_1}
\eta^{a}_{\mu} +...,\\
\tilde{\delta}_{\eta} A^{a}_{\mu\lambda_1 \lambda_2} &=& \partial_{\lambda_1}
\eta^{a}_{\mu\lambda_2} + \partial_{\lambda_2}
\eta^{a}_{\mu\lambda_1} +...\nn
\eeqa
This symmetry appears in addition to the extended
gauge transformations $\delta_{\xi}$ (\ref{polygauge}).
Two families of tensor gauge parameters $\{  \xi \}$ and $\{ \eta \}$
have a common Yang-Mills subgroup which is described by the scalar parameters
$\xi^{a} \equiv \eta^{a}$ .
It is instructive to compare these transformations.
The transformations $\delta_{\xi}$ and $\tilde{\delta}_{\eta}$ do not
coincide and are complementary to each other in the following sense:
in $\delta_{\xi}$
the derivatives of the gauge parameters $\{  \xi \}$ are  over the first
index $\mu$, while in $\tilde{\delta}_{\eta}$ the derivatives of the
gauge parameters $\{ \eta \}$ are over the rest of totally symmetric
indices $\lambda_1 ... \lambda_{s}$ so that together they cover all
indices of the nonsymmetric tensor gauge fields $
A^{a}_{\mu\lambda_1 ... \lambda_{s}}(x)$
(recall that these tensor gauge fields  are not symmetric with respect
to the index $\mu$ and the rest of the indices $\lambda_1 ... \lambda_{s}$).

If one considers the sum of complementary gauge
transformations $\delta_{\xi} +\tilde{\delta}_{\xi} $ acting on
{\it free and totally symmetric  } Abelian tensor gauge fields
then one can
find that it is equivalent to a gauge transformation defined
in the literature \cite{fierz,fierzpauli,yukawa1,wigner,schwinger,
Weinberg:1964cn,chang,singh,fronsdal,deWit:1979pe}, but without
any restrictions on the gauge parameters.
They are also in the same spirit as the gauge transformation
of {\it free} Abelian tensor gauge fields
with "mixed symmetries" considered in \cite{Curtright:1980yk}.
The tensor gauge fields $A^{a}_{\mu\lambda_1 ... \lambda_{s}}(x)$
appear to be more general because their index permutation
symmetry does not correspond to any given Young diagram.

For non-zero values of the coupling constant $g$ the
full transformation $\delta_{\xi}$ (\ref{polygauge})
has the following form \cite{Savvidy:2005fi,Savvidy:2005zm,Savvidy:2005ki}:
\beqa\label{polygaugefull}
\delta A^{a}_{\mu} &=& ( \delta^{ab}\partial_{\mu}
+g f^{acb}A^{c}_{\mu})\xi^b ,~~~~~\\
\delta A^{a}_{\mu\nu} &=&  ( \delta^{ab}\partial_{\mu}
+  g f^{acb}A^{c}_{\mu})\xi^{b}_{\nu} + g f^{acb}A^{c}_{\mu\nu}\xi^{b},\nonumber\\
\delta A^{a}_{\mu\nu \lambda}& =&  ( \delta^{ab}\partial_{\mu}
+g f^{acb} A^{c}_{\mu})\xi^{b}_{\nu\lambda} +
g f^{acb}(  A^{c}_{\mu  \nu}\xi^{b}_{\lambda } +
A^{c}_{\mu \lambda }\xi^{b}_{ \nu}+
A^{c}_{\mu\nu\lambda}\xi^{b}),\nn\\
.........&.&............................\nn
\eeqa
It was important to know the complementary gauge transformation (\ref{freedoublepolygaugesymmetric}) for non-zero values of the coupling constant $g$ as well. It appears that
its unique form can be fixed by the requirement that $\tilde{\delta}_{\eta}$
should form a  group, and the full transformation
(\ref{freedoublepolygaugesymmetric})  takes the following form
\cite{Savvidy:2005ki} :
\beqa\label{doublepolygaugesymmetric}
\tilde{\delta}_{\eta} A^{a}_{\mu} &=& ( \delta^{ab}\partial_{\mu}
+g f^{acb}A^{c}_{\mu})\eta^b ,\\
\tilde{\delta}_{\eta} A^{a}_{\mu\lambda_1} &=&  ( \delta^{ab}\partial_{\lambda_1}
+  g f^{acb}A^{c}_{\lambda_1})\eta^{b}_{\mu} + g f^{acb}A^{c}_{\mu\lambda_1}\eta^{b},\nn\\
\tilde{\delta}_{\eta} A^{a}_{\mu\lambda_1\lambda_2} &=& ( \delta^{ab}\partial_{\lambda_1}
+g f^{acb} A^{c}_{\lambda_1})\eta^{b}_{\mu\lambda_2} +( \delta^{ab}\partial_{\lambda_2}
+g f^{acb} A^{c}_{\lambda_2})\eta^{b}_{\mu\lambda_1} +\nn\\
&~&
+g f^{acb}(  A^{c}_{\mu  \lambda_1}\eta^{b}_{\lambda_2 }+
A^{c}_{\mu \lambda_2 }\eta^{b}_{\lambda_1}+
A^{c}_{\lambda_1\lambda_2}\eta^{b}_{\mu} +
A^{c}_{\lambda_2 \lambda_1}\eta^{b}_{\mu} +A^{c}_{\mu\lambda_1\lambda_2}\eta^{b}).\nn
\eeqa
It forms a closed algebraic structure (see the last section and Appendix A )
\be\label{gaugecommutator}
[~\tilde{\delta}_{\eta},\tilde{\delta}_{\chi}]~A_{\mu\lambda_1\lambda_2 ...\lambda_s} ~=~
-i g~ \tilde{\delta}_{\zeta} A_{\mu\lambda_1\lambda_2 ...\lambda_s}
\ee
with the same composition law for the gauge parameters as for the transformation
$\delta_{\xi}$:
\beqa\label{gaugealgebra}
\zeta&=&[\eta,\chi]\\
\zeta_{\lambda_1}&=&[\eta,\chi_{\lambda_1}] +[\eta_{\lambda_1},\chi]\nn\\
\zeta_{\lambda_1\lambda_2} &=& [\eta,\chi_{\lambda_1\lambda_2}] +
[\eta_{\lambda_1},\chi_{\lambda_2}]
+ [\eta_{\lambda_2},\chi_{\lambda_1}]+[\eta_{\lambda_1\lambda_2},\chi],\nn\\
......&.&..........................\nn
\eeqa
This means that  (\ref{polygaugefull}) and (\ref{doublepolygaugesymmetric})
can be considered as "complementary"   representations of the same infinite-dimensional
gauge group ${\cal G}$ with algebra (\ref{gaugealgebra})
\cite{Savvidy:2005ki}.

\section{\it Complementary  Field Strength Tensors}

The field strength tensors $G^{a}_{\mu\nu,\lambda_1 ... \lambda_s}$ transform
{\it homogeneously} with respect to the transformations $\delta_{\xi}$
(\ref{polygaugefull})
\cite{Savvidy:2005fi,Savvidy:2005zm}
\beqa\label{transformationfieldstrenghparticular}
\delta_{\xi} G^{a}_{\mu\nu}&=& g f^{abc} G^{b}_{\mu\nu} \xi^c \\
\delta_{\xi} G^{a}_{\mu\nu,\lambda} &=& g f^{abc} (~G^{b}_{\mu\nu,\lambda} \xi^c
+ G^{b}_{\mu\nu} \xi^{c}_{\lambda}~),\nonumber\\
\delta_{\xi} G^{a}_{\mu\nu,\lambda\rho} &=& g f^{abc}
(~G^{b}_{\mu\nu,\lambda\rho} \xi^c
+ G^{b}_{\mu\nu,\lambda} \xi^{c}_{\rho} +
G^{b}_{\mu\nu,\rho} \xi^{c}_{\lambda} +
G^{b}_{\mu\nu} \xi^{c}_{\lambda\rho}~)\nn\\
......&.&..........................,\nn
\eeqa
but {\it inhomogeneously} with respect to the
complementary gauge transformations $\tilde{\delta}_{\eta}$
(\ref{doublepolygaugesymmetric}). The natural question
which arises in this respect is the following: do there exist
"complementary" field strength tensors $\tilde{G}^{a}_{\mu\nu,\lambda_1 ... \lambda_s}$
which transform homogeneously,
now with respect to the $\tilde{\delta}_{\eta}$ ~? And if yes, how can one construct
new invariants~?
The answer to the above questions is affirmative and we shall present the form
of the $\tilde{G}^{a}_{\mu\nu,\lambda}$ and
$\tilde{G}^{a}_{\mu\nu,\lambda \rho}$ and the corresponding invariants.
We shall define  field strength tensors as follows:
\beqa\label{fieldstrengthparticular}
\tilde{G}^{a}_{\mu\nu} &\equiv& G^{a}_{\mu\nu}=
\partial_{\mu} A^{a}_{\nu} - \partial_{\nu} A^{a}_{\mu} +
g f^{abc}~A^{b}_{\mu}~A^{c}_{\nu},\\
\tilde{G}^{a}_{\mu\nu,\lambda} &=&
\partial_{\mu} A^{a}_{\lambda\nu} - \partial_{\nu} A^{a}_{\lambda\mu} +
g f^{abc}(~A^{b}_{\mu}~A^{c}_{\lambda\nu} + A^{b}_{\lambda\mu}~A^{c}_{\nu} ~),
\nn\\
\tilde{G}^{a}_{\mu\nu,\lambda\rho} &=& {1\over 2}\{~
\partial_{\mu}( A^{a}_{\lambda\nu\rho}+A^{a}_{\rho\nu\lambda} -A^{a}_{\nu\lambda\rho})
+ g f^{abc}~A^{b}_{\mu}~
( A^{a}_{\lambda\nu\rho}+A^{a}_{\rho\nu\lambda} -A^{a}_{\nu\lambda\rho}) +\nn\\
&-& \partial_{\nu} ( A^{a}_{\lambda\mu\rho} + A^{a}_{\rho\mu\lambda}- A^{a}_{\mu\lambda\rho}) +
g f^{abc}~
( A^{a}_{\lambda\mu\rho} + A^{a}_{\rho\mu\lambda}-
A^{a}_{\mu\lambda\rho})~A^{c}_{\nu} ~\}
\nn\\
&+& g f^{abc}~(~A^{b}_{\lambda\mu}~A^{c}_{\rho\nu}+A^{b}_{\rho\mu}~A^{c}_{\lambda\nu}~).\nn
\eeqa
The complementary field strength tensors are
antisymmetric in their first two indices and are totally symmetric
with respect to the rest of the indices. The symmetry properties of the field strength tensors
$\tilde{G}^{a}_{\mu\nu,\lambda}$ and $\tilde{G}^{a}_{\mu\nu,\lambda \rho}$
remain invariant in the course of this transformation.
As one can show by
direct computation, they transform {\it homogeneously} with
respect to the complementary
gauge transformations $\tilde{\delta}_{\eta}$ (\ref{doublepolygaugesymmetric})
\footnote{See the next chapter for the derivation of these formulas.}
\beqa\label{fieldstrenghparticulartransformation}
\tilde{\delta}_{\eta} G^{a}_{\mu\nu}&=& g f^{abc} G^{b}_{\mu\nu} \eta^c \\
\tilde{\delta}_{\eta} \tilde{G}^{a}_{\mu\nu,\lambda} &=& g f^{abc} (~\tilde{G}^{b}_{\mu\nu,\lambda} \eta^c
+ G^{b}_{\mu\nu} \eta^{c}_{\lambda}~),\nonumber\\
\tilde{\delta}_{\eta} \tilde{G}^{a}_{\mu\nu,\lambda\rho} &=& g f^{abc}
(~\tilde{G}^{b}_{\mu\nu,\lambda\rho} \eta^c
+ \tilde{G}^{b}_{\mu\nu,\lambda} \eta^{c}_{\rho} +
\tilde{G}^{b}_{\mu\nu,\rho} \eta^{c}_{\lambda} +
G^{b}_{\mu\nu} \eta^{c}_{\lambda\rho}~).\nn
\eeqa
The form of these transformations
is identical with the one for  the field strength tensors
$\delta_{\xi}G^{a}_{\mu\nu,\lambda_1 ... \lambda_s}$
given by the formulae (\ref{transformationfieldstrenghparticular}).
This simply means that the invariant forms can be constructed in
the same way as for the transformation
$\delta_{\xi}$ in \cite{Savvidy:2005fi,Savvidy:2005zm}. They are
$\tilde{{{\cal L}}}_{2}$ and $\tilde{{{\cal L}}}^{'}_{2}$
and are quadratic in   $\tilde{G}^{a}_{\mu\nu,\lambda_1 ... \lambda_s}$:
\beqa\label{alternativetotalactiontwo}
\tilde{{{\cal L}}}(A)  = {{\cal L}}_1 +
g_2 (\tilde{{{\cal L}}}_2 +  \tilde{{{\cal L}}}^{'}_2 )=
&-&{1\over 4}G^{a}_{\mu\nu}G^{a}_{\mu\nu} +\\
+ g_2 \{&-&{1\over 4}\tilde{G}^{a}_{\mu\nu,\lambda}\tilde{G}^{a}_{\mu\nu,\lambda}
-{1\over 4}G^{a}_{\mu\nu}\tilde{G}^{a}_{\mu\nu,\lambda\lambda} +\nn\\
&+&{1\over 4}\tilde{G}^{a}_{\mu\nu,\lambda}\tilde{G}^{a}_{\mu\lambda,\nu}
+{1\over 4}\tilde{G}^{a}_{\mu\nu,\nu}\tilde{G}^{a}_{\mu\lambda,\lambda}
+{1\over 2}G^{a}_{\mu\nu}\tilde{G}^{a}_{\mu\lambda,\nu\lambda} \}\nn
\eeqa
Thus we have two Lagrangian forms ${{\cal L}}(A)$ in (\ref{lagrangian}) and
$\tilde{{\cal L}}(A)$ in (\ref{alternativetotalactiontwo})
for the same lower-rank tensor gauge fields.
They are fully invariant with respect
to the corresponding gauge transformations (\ref{polygaugefull}) and
(\ref{doublepolygaugesymmetric})
\be
\delta_{\xi}{{\cal L}}(A)=0, ~~~~~~~\tilde{\delta}_{\eta}\tilde{{\cal L}}(A)=0.
\ee
The natural question which arises at this point is to find out a
possible relation between these Lagrangian forms.

First of all one can see that the definition of the field strength tensors
$\tilde{G}^{a}_{\mu\nu,\lambda}$ and $\tilde{G}^{a}_{\mu\nu,\lambda \rho}$
in (\ref{fieldstrengthparticular})  is the same as for the field
strength tensors $G^{a}_{\mu\nu,\lambda}$ and $G^{a}_{\mu\nu,\lambda \rho}$,
if one defines  dual fields  as follows:
\beqa\label{dualitytransformation}
  \begin{array}{ll}
\tilde{A}_{\mu\nu} =  A_{\nu\mu}   ,  \\
\tilde{A}_{\mu\nu\lambda} =
{1\over 2}(A_{\lambda\mu\nu} + A_{\nu\mu\lambda} - A_{\mu\nu\lambda}) .
 \end{array}
\eeqa
Then
\beqa\label{dualfields}
\tilde{G}_{\mu\nu,\lambda}(A)  &=& G_{\mu\nu,\lambda}(\tilde{A}),\nn\\
\tilde{G}_{\mu\nu,\lambda\rho}(A)  &=& G_{\mu\nu,\lambda\rho}(\tilde{A})
\eeqa
and the Lagrangian $\tilde{{\cal L}}(A)$ in (\ref{alternativetotalactiontwo})
is mapped  into the
Lagrangian ${{\cal L}}(\tilde{A})$ in (\ref{lagrangian})
\be\label{dualmap}
\tilde{{{\cal L}}}(A) ~~\rightarrow ~~{{\cal L}}(\tilde{A}).
\ee
Therefore the above transformation
(\ref{dualitytransformation}) can be considered as a duality transformation
which  allows us to map  the Lagrangian $\tilde{{\cal L}}$ into the Lagrangian
$ {\cal L}$.
One can also define the inverse dual transformation as
\be
 \begin{array}{ll}
A_{\nu\mu} =  \tilde{A}_{\mu\nu}   ,  \\
A_{\mu\nu\lambda} =   \tilde{A}_{\lambda\mu\nu} + \tilde{A}_{\nu \mu\lambda}.
 \end{array}
\ee
It has the property that $\tilde{A}(A(\tilde{A})) = \tilde{A}$ and $A(\tilde{A}(A))= A$
and therefore the dual map  is one-to-one.

\section{\it Gauge Transformation of Field Strength Tensors}
We shall compute here the variation of the field strength tensors
$\tilde{G}^{a}_{\mu\nu,\lambda}$ and $\tilde{G}^{a}_{\mu\nu,\lambda \rho}$
under the complementary gauge transformation
$\tilde{\delta}_{\eta}$ (\ref{doublepolygaugesymmetric}) in matrix form.
We have
\beqa\label{varationfieldstrengh}
\tilde{\delta}_{\eta} \tilde{G}_{\mu\nu,\lambda}&=&\partial_{\mu}
\{ \partial_{\nu}\eta_{\lambda} -i g [A_\nu , \eta_{\lambda}]
-i g [A_{\lambda\nu} , \eta]  \}
-\partial_{\nu}\{ \partial_{\mu} \eta_{\lambda} -i g [A_\mu , \eta_{\lambda}]
-i g [A_{\lambda\mu} , \eta]  \}
\nn\\
&-&i g [ \partial_{\mu} \eta -i g [A_\mu , \eta ],  A_{\lambda\nu}]
- i g [A_\mu , \partial_{\nu}\eta_{\lambda} -i g [A_\nu , \eta_{\lambda}]
-i g [A_{\lambda\nu} , \eta]]\nn\\
&-&i g [A_{\lambda\mu}, \partial_{\nu} \eta -i g [A_\nu , \eta ] ]
- i g [\partial_{\mu}\eta_{\lambda} -i g [A_\mu , \eta_{\lambda}]
-i g [A_{\lambda\mu} , \eta], A_\nu]
\\
&=& -i g [ \partial_{\mu}A_{\nu}  - \partial_{\nu}A_{\mu}
-i g [A_\mu , A_{\nu}],\eta_{\lambda}] -\nn\\
&& - i g [ \partial_{\mu}A_{\lambda\nu}  - \partial_{\nu}A_{\lambda\mu}
-i g [A_\mu , A_{\lambda\nu}] -i g [A_{\lambda\mu} , A_{\nu}] ,\eta] =\nn\\
&=& -i g [ \tilde{G}_{\mu\nu ,\lambda}~, \eta]
 - i g [ G_{\mu  \nu} ~ ,\eta_{\lambda}] \nn
\eeqa
and for the $\tilde{\delta}_{\eta} \tilde{G}^{a}_{\mu\nu,\lambda\rho}$ we get
\beqa
\tilde{\delta}_{\eta} \tilde{G}^{a}_{\mu\nu,\lambda\rho}&=&\partial_{\mu}
\{ \partial_{\nu}\eta_{\lambda\rho} -i g [A_\nu , \eta_{\lambda\rho}]
-i g [A_{\lambda\nu} , \eta_{\rho}] -i g [A_{\rho\nu} , \eta_{\lambda}] \}
\\
&-& i g {1\over 2}\partial_{\mu}[A_{\lambda\nu\rho} +A_{\rho\nu\lambda}-
A_{\nu\lambda\rho},\eta]\nn\\
&-&\partial_{\nu}\{ \partial_{\mu} \eta_{\lambda\rho} -i g [A_\mu , \eta_{\lambda\rho}]
-i g [A_{\lambda\mu} , \eta_{\rho}]  -i g [A_{\rho\mu} , \eta_{\lambda}] \}\nn\\
&-& i g {1\over 2}\partial_{\nu}[A_{\lambda\mu\rho} +A_{\rho\mu\lambda}-
A_{\mu\lambda\rho},\eta]
\nn\\
&-&i g {1\over 2}[ \partial_{\mu} \eta -i g [A_\mu , \eta ],
A_{\lambda\nu\rho} +A_{\rho\nu\lambda}-A_{\nu\lambda\rho}]\nn\\
&-& i g [A_\mu , \partial_{\nu}\eta_{\lambda\rho} -i g [A_\nu , \eta_{\lambda\rho}]
-i g [A_{\lambda\nu} , \eta_{\rho}] -i g [A_{\rho\nu} , \eta_{\lambda}]
-{1\over 2} [A_{\lambda\nu\rho} +A_{\rho\nu\lambda}-A_{\nu\lambda\rho}, \eta ]]
\nn\\
&-&i g {1\over 2}[A_{\lambda\mu\rho} +A_{\rho\mu\lambda}-A_{\mu\lambda\rho} ,
\partial_{\nu} \eta -i g [A_\nu , \eta ]]
\nn\\
&-& i g [\partial_{\mu}\eta_{\lambda\rho} -i g [A_\mu , \eta_{\lambda\rho}]
-i g [A_{\lambda\mu} , \eta_{\rho}] -i g [A_{\rho\mu} , \eta_{\lambda}]
-{1\over 2} [A_{\lambda\mu\rho} +A_{\rho\mu\lambda}-A_{\mu\lambda\rho}, \eta ],A_\nu]
\nn\\
&-& i g [\partial_{\mu}\eta_{\lambda} -i g [A_\mu , \eta_{\lambda}]
-i g [A_{\lambda\mu} , \eta], A_{\rho\nu}]
-i g [A_{\lambda\mu}, \partial_{\nu} \eta_{\rho} -i g [A_\nu , \eta_{\rho}]
-i g [A_{\rho\nu},  \eta]]
\nn\\
&-& i g [\partial_{\mu}\eta_{\rho} -i g [A_\mu , \eta_{\rho}]
-i g [A_{\rho\mu} , \eta], A_{\lambda\nu}]
-i g [A_{\rho\mu}, \partial_{\nu} \eta_{\lambda} -i g [A_\nu , \eta_{\lambda}]
-i g [A_{\lambda\nu},  \eta]]=
\nn\\
&=& -i g
[~\tilde{G}_{\mu\nu,\lambda\rho} ~,\eta]
+ [\tilde{G}_{\mu\nu,\lambda} ~,\eta_{\rho}] +
[\tilde{G}_{\mu\nu,\rho} ~,\eta_{\lambda}] +
[G_{\mu\nu} ~,\eta_{\lambda\rho}~]. \nn
\eeqa
and we arrive at the  result  (\ref{fieldstrenghparticulartransformation}).

\section{\it Extension to High-Rank Tensors}

It is important to find out the complementary gauge transformation
$\tilde{\delta}_{\eta}$ acting on higher-rank tensor gauge fields.
This transformation was known up to the tensor gauge
fields of rank three  and
was presented above by the formula (\ref{doublepolygaugesymmetric})
\cite{Savvidy:2005ki}.
Below we shall present the $\tilde{\delta}_{\eta}$ transformation acting
on a rank-4 gauge field. It is presented
in a matrix form because it is much easier to use for algebraic calculations.
The transformation is:
\beqa\label{dualpolygaugetransformation}
\tilde{\delta}_{\eta} A_{\mu\lambda_1\lambda_2 \lambda_3} &=&
\nabla_{\lambda_1}
\eta_{\mu\lambda_2\lambda_3} +\nabla_{\lambda_2}\eta_{\mu\lambda_3\lambda_1}
+\nabla_{\lambda_3}\eta_{\mu\lambda_1\lambda_2}-\nn\\
&-&i g [A_{\mu  \lambda_1} , \eta_{\lambda_2 \lambda_3}]
-i g [A_{\mu \lambda_2 } ,\eta_{\lambda_3\lambda_1}]
-i g [A_{\mu \lambda_3 } ,\eta_{\lambda_1\lambda_2}] -\nn\\
&-&i g [A_{\lambda_1\lambda_2} +A_{\lambda_2 \lambda_1}, \eta_{\mu\lambda_3}]
-i g [A_{\lambda_1\lambda_3} +A_{\lambda_3 \lambda_1}, \eta_{\mu\lambda_2}]
-i g [A_{\lambda_2\lambda_3} +A_{\lambda_3 \lambda_2}, \eta_{\mu\lambda_1}]
-\nn\\
&-&i g [A_{\mu\lambda_1\lambda_2} ,\eta_{\lambda_3}]
-i g [A_{\mu\lambda_1\lambda_3} ,\eta_{\lambda_2}]
-i g [A_{\mu\lambda_2\lambda_3} ,\eta_{\lambda_1}] - \\
&-&{1\over 2}i g [A_{\lambda_1\lambda_2\lambda_3}
+A_{\lambda_2\lambda_3\lambda_1} +A_{\lambda_3\lambda_1\lambda_2},\eta_{\mu}]
-i g [A_{\mu\lambda_1\lambda_2\lambda_3} , \eta ], \nn
\eeqa
and should be considered together with (\ref{doublepolygaugesymmetric}).
The corresponding field strength tensor is defined by the formula
\beqa\label{fieldstrengthparticularextens}
\tilde{G}_{\mu\nu,\lambda_1 \lambda_2\lambda_3} &=&
\nabla_{\mu} \{{1\over 3}(A_{\lambda_1\nu\lambda_2\lambda_3}
+ A_{\lambda_2\nu\lambda_1\lambda_3}
+ A_{\lambda_3\nu\lambda_1\lambda_2})
-{2\over 3} A_{\mu\lambda_1\lambda_2 \lambda_3 }\}
 \\
&-&\nabla_{\nu}  \{{1\over 3}(A_{\lambda_1\mu\lambda_2\lambda_3}
+ A_{\lambda_2\mu\lambda_1\lambda_3}
+ A_{\lambda_3\mu\lambda_1\lambda_2})
-{2\over 3} A_{\mu\lambda_1\lambda_2 \lambda_3 }\}
\nn\\
&-& i g [ A_{\lambda_1\mu}, ~{1\over 2}(A_{\lambda_2\nu\lambda_3} +
A_{\lambda_3\nu\lambda_2})
-{1\over 2} A_{\nu\lambda_2\lambda_3}]
\nn\\
&-& i g [ A_{\lambda_2\mu},~ {1\over 2}(A_{\lambda_1\nu\lambda_3} +
A_{\lambda_3\nu\lambda_1})
-{1\over 2} A_{\nu\lambda_1\lambda_3}]
\nn\\
&-& i g [ A_{\lambda_3\mu}, ~{1\over 2}(A_{\lambda_1\nu\lambda_2} +
A_{\lambda_2\nu\lambda_1})
-{1\over 2} A_{\nu\lambda_1\lambda_2}]-
\nn\\
&-&i g [{1\over 2}(A_{\lambda_1\mu\lambda_2} + A_{\lambda_2\mu\lambda_1})
-{1\over 2} A_{\mu\lambda_1\lambda_2},~ A_{\lambda_3\nu}]
\nn\\
&-&i g [{1\over 2}(A_{\lambda_1\mu\lambda_3} + A_{\lambda_3\mu\lambda_1})
-{1\over 2} A_{\mu\lambda_1\lambda_3}, ~A_{\lambda_2\nu }]
\nn\\
&-&i g [{1\over 2}(A_{\lambda_2\mu\lambda_3} + A_{\lambda_2\mu\lambda_3})
-{1\over 2} A_{\mu\lambda_2\lambda_3}, ~A_{\lambda_1\nu}]\nn
\eeqa
and transforms homogeneously. The duality transformation
(\ref{dualitytransformation}) will take the form
\beqa\label{dualityfieldtransformations}
  \begin{array}{ll}
\tilde{A}_{\mu\lambda_1} =  A_{\lambda_1\mu}   ,  \\
\tilde{A}_{\mu\lambda_1\lambda_2} =
{1\over 2}(A_{\lambda_1\mu\lambda_2} + A_{\lambda_2\mu\lambda_1})
-{1\over 2} A_{\mu\lambda_1\lambda_2} \\
\tilde{A}_{\mu\lambda_1\lambda_2\lambda_3} =
{1\over 3}(A_{\lambda_1\mu\lambda_2\lambda_3}
+ A_{\lambda_2\mu\lambda_1\lambda_3}
+ A_{\lambda_3\mu\lambda_1\lambda_2})
-{2\over 3} A_{\mu\lambda_1\lambda_2 \lambda_3 } \\
 \end{array}
\eeqa
and tells us that
\beqa
\tilde{G}_{\mu\nu,\lambda_1\lambda_2}(A)  &=&
G_{\mu\nu,\lambda_1\lambda_2}(\tilde{A}).
\eeqa
It is a natural extension of the transformation
(\ref{dualfields}) and most probably will extend to all,
properly defined, higher-rank complementary field strength tensors
$$
\tilde{G}_{\mu\nu,\lambda_1 ... \lambda_s}(A)  =
G_{\mu\nu,\lambda_1 ... \lambda_s}(\tilde{A})
$$
where the dual fields (\ref{dualitytransformation}),
(\ref{dualityfieldtransformations}) are defined as follows
\be
\tilde{A}_{\mu\lambda_1 ...  \lambda_s} =
{1\over s}(A_{\lambda_1\mu...\lambda_s}
+ ....
+ A_{\lambda_s\mu \lambda_{s-1}})
-{s-1\over s} A_{\mu\lambda_1 ... \lambda_s }~~~~s=1,2,.....
\ee
Therefore it seems that we shall have the duality map also for
the higher-rank invariants
$$
\tilde{g}_s \tilde{{\cal L}}_{s}~ + ~\tilde{g}^{'}_s
\tilde{{\cal L}}^{'}_{s}~~~\rightarrow~~~
g_s{{\cal L}}_{s}~ +~ g^{'}_s {{\cal L}}^{'}_{s}   .
$$
We shall leave this extension for the future studies.

It is a great pleasure to express our thanks to Thordur Jonsson for
stimulating discussions and his kind hospitality of one of us (G.S.) in the
University of Iceland where part of this work was completed. The work of
(J.K.B.) was supported by the Icelandic Research Fund.  One of us
(G.S.) would like to thank Takuya Tsukioka 
for pointing out the mistake in the gauge transformation of
$\tilde{\delta}  A_{\mu\lambda_1\lambda_2}$.
This work was partially supported by the EEC Grant no. MRTN-CT-2004-005616.

\section{\it Appendix A}
Let us prove that a commutator of two $\tilde{\delta}_{\eta}$ transformations
can be expressed as a similar gauge transformation, and therefore   gauge
transformations (\ref{doublepolygaugesymmetric}) form a closed algebraic structure.
To make the calculation more transparent let us
express the transformation law (\ref{doublepolygaugesymmetric}) in a matrix form:
\beqa\label{matrixform}
\tilde{\delta}_{\eta}  A_{\mu} &=& \partial_{\mu}\eta -i g[A_{\mu},\eta]\nonumber\\
\tilde{\delta}_{\eta}  A_{\mu\nu} &=& \partial_{\nu}\eta_{\mu} -i g[A_{\nu},\eta_{\mu}]
-i g [A_{\mu\nu},\eta]\nonumber\\
\tilde{\delta}_{\eta}  A_{\mu\nu\lambda} &=& \partial_{\nu}\eta_{\mu\lambda}
-i g[A_{\nu},\eta_{\mu\lambda}] +\partial_{\lambda}\eta_{\mu\nu}
-i g[A_{\lambda},\eta_{\mu\nu}]-\nn\\
&-&i g[A_{\mu\nu},\eta_{\lambda}]-i g [A_{\mu\lambda},\eta_{\nu}]-
i g[A_{\lambda\nu},\eta_{\mu}]-i g [A_{\nu\lambda},\eta_{\mu}]
-i g [A_{\mu\nu\lambda},\eta],
\eeqa
where $A_{\mu\nu} = A^{a}_{\mu\nu} L^a$, $A_{\mu\nu\lambda} =
A^{a}_{\mu\nu\lambda} L^a$ and $\xi = L^{a} \xi^{a}$.
The commutator of two gauge transformations acting
on a second-rank tensor gauge field is:
\beqa
[\tilde{\delta}_{\eta},\tilde{\delta}_{\chi}]A_{\mu\nu} &=&
\tilde{\delta}_{\eta}~(-i g[A_{\nu},\chi_{\mu}]
-i g [A_{\mu\nu},\chi])-\nonumber\\
&-& \tilde{\delta}_{\chi}~(-i g[A_{\nu},\eta_{\mu}]
-i g [A_{\mu\nu},\eta])\nonumber\\
&=&-ig~\{~\partial_{\nu}([\eta,\chi_{\mu}] +[\eta_{\mu},\chi]) -i
g[A_{\nu},([\eta,\chi_{\mu}] +[\eta_{\mu},\chi])]
-i g[A_{\mu\nu},[\eta,\chi]] ~\}\nonumber\\
&=& -ig~\{~\partial_{\nu}\zeta_{\mu} -i g[A_{\nu},\zeta_{\mu}]
-i g [A_{\mu\nu},\zeta]~\} = -i g ~\tilde{\delta}_{\zeta}A_{\mu\nu}\nn
\eeqa
and is again a gauge transformation with gauge parameters
$\zeta^a, \zeta^{a}_{\mu}$ which are given by the following expressions:
$$
\zeta =[\eta,\chi],~~~~~~~\zeta_{\nu} = [\eta,\chi_{\nu}] +[\eta_{\nu},\chi].
$$
The commutator of two gauge transformations acting on a rank-3 tensor gauge field is:
\beqa
[\tilde{\delta}_{\eta},\tilde{\delta}_{\chi}]A_{\mu\nu\lambda} &=&
\tilde{\delta}_{\eta}~(-i g[A_{\nu},\chi_{\mu\lambda}]
-i g[A_{\lambda},\chi_{\mu\nu}]
-i g  [A_{\mu\nu},\chi_{\lambda}]-i g  [A_{\mu\lambda},\chi_{\nu}]-\nn\\
&-&i g  [A_{\nu\lambda},\chi_{\mu}]-i g  [A_{\lambda\nu},\chi_{\mu}]-
i g [A_{\mu\nu\lambda},\chi]) -
\nn\\
&-&\tilde{\delta}_{\chi}~(-i g[A_{\nu},\eta_{\mu\lambda}]
-i g[A_{\lambda},\eta_{\mu\nu}]
-i g  [A_{\mu\nu},\eta_{\lambda}]-i g  [A_{\mu\lambda},\eta_{\nu}]-\nn\\
&-&i g  [A_{\nu\lambda},\eta_{\mu}]-i g  [A_{\lambda\nu},\eta_{\mu}]-
i g [A_{\mu\nu\lambda},\eta])
\nonumber\\
&=&-ig\{~\partial_{\nu} ( [\eta,\chi_{\mu\lambda}] + [\eta_{\mu},\chi_{\lambda}]
+ [\eta_{\lambda},\chi_{\mu}] + [\eta_{\mu\lambda},\eta])\nonumber\\
&~&~-ig [A_{\nu},( [\eta,\chi_{\mu\lambda}] + [\eta_{\mu},\chi_{\lambda}]
+ [\eta_{\lambda},\chi_{\mu}] + [\eta_{\mu\lambda},\eta])]\nonumber\\
&&+\partial_{\lambda} ( [\eta,\chi_{\mu\nu}] + [\eta_{\mu},\chi_{\nu}]
+ [\eta_{\nu},\chi_{\mu}] + [\eta_{\mu\nu},\eta])\nonumber\\
&~&~-ig [A_{\lambda},( [\eta,\chi_{\mu\nu}] + [\eta_{\mu},\chi_{\nu}]
+ [\eta_{\nu},\chi_{\mu}] + [\eta_{\mu\nu},\eta])]
\nonumber\\
&~&~-i g  [A_{\mu\nu},( [\eta,\chi_{\lambda}]  + [\eta_{\lambda},\chi])]
-i g  [A_{\mu\lambda},( [\eta,\chi_{\nu}]  + [\eta_{\nu},\chi])]-\nn\\
&~&~-i g  [A_{ \nu\lambda},( [\eta,\chi_{\mu}]  + [\eta_{\mu},\chi])]
-i g  [A_{ \lambda\nu},( [\eta,\chi_{\mu}]  + [\eta_{\mu},\chi])]-
i g[A_{\mu\nu\lambda},[\eta,\chi] ] \}\nonumber\\
&=&-ig~\{~\partial_{\nu} \zeta_{\mu\lambda} -i g[A_{\nu},\zeta_{\mu\lambda}]+
~\partial_{\lambda} \zeta_{\mu\nu} -i g[A_{\lambda},\zeta_{\mu\nu}] -
i g  [A_{\mu\nu},\zeta_{\lambda}]-i g   [A_{\mu\lambda},\zeta_{\nu}]-\nn\\
&-&i g  [A_{\nu\lambda},\zeta_{\mu}]-i g   [A_{\lambda\nu},\zeta_{\mu}]-
i g [A_{\mu\nu\lambda},\zeta]~\} =\tilde{\delta}_{\zeta}A_{\mu\nu\lambda},\nonumber
\eeqa
where
\beqa\label{commutatorofparameterslow}
\zeta =[\eta,\xi],~~~\zeta_{\nu} = [\eta,\xi_{\nu}] +[\eta_{\nu},\xi],~~~~
\zeta_{\nu\lambda} = [\eta,\xi_{\nu\lambda}] +  [\eta_{\nu},\xi_{\lambda}]
+ [\eta_{\lambda},\xi_{\nu}]+[\eta_{\nu\lambda},\xi].
\eeqa
It is also instructive to consider the transformation properties of the
dual field $\tilde{A}_{\mu\nu\lambda}$ in (\ref{dualitytransformation})
under the transformations $\tilde{\delta}_\eta$.
It takes the following form
\be
\tilde{\delta}_\eta  \tilde{A}_{\mu\nu\lambda} = \partial_{\mu}\eta_{\nu\lambda}
-i g[A_{\mu},\eta_{\nu\lambda}]-
i g[\tilde{A}_{\mu\nu},\eta_{\lambda}]-i g [\tilde{A}_{\mu\lambda},\eta_{\nu}]
-i g [\tilde{A}_{\mu\nu\lambda},\eta]
\ee
and coincides with the transformation law
$\delta_{\xi}A_{\mu\nu\lambda}$ (\ref{polygaugefull}).

\end{document}